\documentclass[twocolumn,amsmath,amssymb,pra,showpacs]{revtex4}
\usepackage{setspace}
\usepackage{graphicx}
\usepackage{amsmath}
\usepackage{amssymb}
\usepackage{latexsym}

\usepackage{natbib}

\begin{document}

\title{The dynamics of dark solitons in a trapped superfluid Fermi gas}
\author{R.G. Scott$^1$,  F. Dalfovo$^1$, L.P. Pitaevskii$^{1,2}$, S. Stringari$^1$}
\affiliation{$^1$INO-CNR BEC Center and Dipartimento di Fisica, Universit{\`a} di Trento, Via Sommarive 14, I-38123 Povo, Italy.\\$^{2}$Kapitza institute for physical problems, ul. Kosygina 2, 119334 Moscow, Russia.} 
\date{10/11/10}



\begin{abstract}
We study soliton oscillations in a trapped superfluid Fermi gas across the Bose-Einstein condensate to Bardeen-Cooper-Schrieffer (BEC-BCS) crossover. We derive an exact equation for the oscillation period in terms of observable quantities, which we confirm by solving the time-dependent Bogoliubov-de Gennes equations. Hence we reveal the appearance and dynamics of solitons across the crossover, and show that the period dramatically increases as the soliton becomes shallower on the BCS side of the resonance. Finally, we propose an experimental protocol to test our predictions.
\end{abstract}

\maketitle

{Solitons have been the focus of much recent research in the field of cold atoms, due to their ubiquitous production in the dynamics of BECs~\cite{kev}. Their different forms create a broad family, from the common ``grey'' and ``black'' solitons in repulsive BECs, to the ``bright'' solitons in attractive BECs and ``gap'' solitons in optical lattices, and their more exotic cousins such as the ``bright-dark'' solitons, which were recently observed in two-component BECs~\cite{Busch,BeckerSH}. We expect solitons to play an equally important role in the dynamics of degenerate Fermi gases. Even more fundamentally, topological excitations offer insights into the nature of coherence and superfluidity across the BEC-BCS crossover, as illustrated by the recent observation of vortex lattices~\cite{KetterleSH}. Despite this interest, the nature of soliton dynamics in Fermi gases remains elusive, and only stationary ``black'' solitons have been simulated across the BEC-BCS crossover~\cite{Antezza}.

In this Letter, we investigate the oscillation of solitons in a harmonic trap. From fundamental statements about the nature of the soliton and the media it moves in, we derive universal relations, valid for both Bosonic and Fermionic superfluids, relating the soliton energy and oscillation period $T_s$ to observable quantities such as phase jump, speed and density. The special case of unitarity is particularly interesting because the soliton width is of the order of interatomic distances and its observation will give access to the short-range physics. We then perform numerical simulations of the time-dependent Bogoliubov-de Gennes (TDBdG) equations~\cite{Challis}. By extracting the appropriate observable quantities from our simulations we find good agreement between the numerical and analytic models, which show that $T_s$ increases dramatically as the soliton becomes shallower when moving from a BEC to a BCS regime. Finally, we propose and simulate an experimental protocol to demonstrate the variation in $T_s$ across the BEC-BCS crossover.

{\it General theory.} Let us consider a soliton in a superfluid gas, which may be either Bosonic or Fermionic, confined in an elongated trap with axial potential $U(x) = m \omega_x^2 x^2 /2$, in which $m$ is the atomic mass and $\omega_x$ is the angular frequency. We assume that the width of the cloud in the axial direction is large compared to the size of the soliton. Then, in a local density approximation, the soliton can be treated as a point-like particle at position $X$; its dynamics can be formulated in terms of the soliton energy in a uniform fluid $E_s\left(\mu,V^2\right)$, where $\mu$ is the chemical potential of the fluid and $V = dX/dt$ is the soliton velocity. Furthermore, we may say that $\mu\left(X\right) = \mu\left(0\right)-U\left(X\right)$. We define the inertial mass $m_I = 2 \partial E_s / \partial V^2$ and the number $N_s = -\partial E_s / \partial \mu = \int^{\infty}_{-\infty}[ n_{1d}(x) - n_{1d0}] \; dx$, in which $n_{1d}(x)$ is the one-dimensional density and $n_{1d0}$ is the bulk density far from the soliton. The quantity $N_s$ is the deficit of particles associated with the depression in density at the soliton position. Note that typically $m_I<0$ and $N_s<0$. Then energy conservation in the absence of dissipation gives~\cite{fedichev,Lev}
\begin{equation}
\frac{dE_s}{dt} = \frac{\partial E_s}{\partial \mu\left(X\right)} \frac{d\mu\left(X\right)}{dX} \frac{dX}{dt} + \frac{\partial E_s}{\partial V^2}\frac{dV^2}{dV}\frac{dV}{dt} = 0
\end{equation}
and hence $m_I(dV/dt)=-N_s(dU/dX) = -N_s m \omega_x^2 X$. For small amplitude oscillations, the soliton period is
\begin{equation}
T_s = \sqrt{m_I/N_s m} \,  T_x ,
\label{eq:freq}
\end{equation}
where $T_x = 2 \pi / \omega_x$, and $m_I$ and $N_s$ are taken for $V\to 0$.

The key to our analytic treatment of solitons is to recognise the distinction between the ``physical'' momentum of the soliton $P_s = m\int^{\infty}_{-\infty} j dx$, associated with the current $j$ carried by the wave function, and the canonical momentum of the soliton $P_c$. By performing a Galilean transformation into the frame of the soliton moving at velocity $V$, we find the current in the soliton frame $\bar{j} = j-n_{1d}V$. Since $\bar{j} = - n_{1d0}V$, we derive
\begin{equation}
P_s = m V \int^{\infty}_{-\infty}[ n_{1d}(x) - n_{1d0}] \; dx = m V N_s.
\label{eq:PS2}
\end{equation}
The canonical momentum of the soliton is instead defined by $V = \partial E_s / \partial P_c$. It follows that 
\begin{equation}
\frac{\partial P_c}{\partial V} = \frac{1}{V}\frac{\partial E_s}{\partial V} = 2 \frac{\partial E_s}{\partial V^2} = m_{I} .
\label{eq:PC2}
\end{equation}

The momenta $P_s$ and $P_c$ are different because, despite being a localized object from the point of view of the density profile and the velocity field, the soliton creates a jump $J_\varphi$ in the phase $\varphi$ of the order parameter. This phase jump is exploited when creating solitons in experiment with the ``phase imprinting'' technique~\cite{BeckerSH, BurgerSH, AndersonSH, DenschlagSH}. In any real experiment, where the soliton is created by a localized perturbation, this phase difference will be compensated by a ``counterflow'', which carries an additional momentum $\Delta P$. The difference between $P_s$ and $P_c$ was first found in a BEC described by the Gross-Pitaevskii (GP) equation~\cite{ishikawa}, and its meaning was discussed in Ref.~\cite{shev}. Far from the soliton we may say that the velocity field is $v = \hbar \nabla\varphi / m_B$, in which $m_B = m$ for Bosons and $m_B = 2m$ for Fermions. Hence we find
$\Delta P = n_{1d0} m \int v dx = \hbar n_{1d0}  J_\varphi m/m_B$, and~\cite{lieblin} 
\begin{equation}
P_c = P_s + \Delta P = P_s + \hbar n_{1d0} J_\varphi m/{m_B} . 
\label{eq:PsPc}
\end{equation} 
Taking into account that $V=0$ for $J_{\varphi}=\pi$, and using Eqs.~(\ref{eq:PS2}) and (\ref{eq:PC2}), we obtain the important relation~\cite{foot2} 
\begin{equation}
mN_{s}V - 2 \int_{0}^{V} \frac{\partial E_s}{\partial V^2} dV
= - \hslash n_{1d0}(J_\varphi - \pi) m/m_{B}  
\; . 
\label{Dphi}
\end{equation}
We then differentiate both sides of Eq.~(\ref{Dphi}) with respect to $V$ to derive
\begin{equation}
m\frac{d\left(N_s V\right)}{dV} - m_{I} = - \frac{\hbar n_{1d0} m}{m_B}\frac{d J_\varphi}{dV} .
\end{equation} 
Substituting $m_I$ using Eq.~(\ref{eq:freq}), for $V \to 0$ we obtain the final result for $T_s$:
\begin{equation}
\left(\frac{T_s}{T_x}\right)^2 - 1 = \frac{\hbar n_{1d0}}{m_B N_s} \frac{d J_\varphi}{dV} \; .
\label{eq:final}
\end{equation}
Note that Eq. (\ref{eq:final}) contains only quantities which can be directly measured in numerical and real experiments. We stress that the basic assumptions used to derive this result are the absence of dissipation and the existence of an order parameter obeying Galilean invariance for particles of mass $m_B$.  

{\it Bogoliubov - de Gennes simulations.} To test Eq.~(\ref{eq:final}), we model the dynamics of a three-dimensional (3D) superfluid Fermi gas by solving the TDBdG equations~\cite{Challis}
\begin{equation}
\left[\begin{array}{ll}
\hat{H} & \Delta \mbox(\textbf{r},t) \\
\Delta^* \mbox(\textbf{r},t) & -\hat{H}
\end{array}\right]
\left[\begin{array}{l}
u_\eta \mbox(\textbf{r},t) \\
v_\eta \mbox(\textbf{r},t)
\end{array}\right] = 
i \hbar \frac{\partial}{\partial t}
\left[\begin{array}{l}
u_\eta \mbox(\textbf{r},t) \\
v_\eta \mbox(\textbf{r},t)
\end{array}\right] , 
\label{eq:tdbdg}
\end{equation}
where $\hat{H} = -\hbar^2 \nabla^2 / 2m + U - \mu\left(0\right)$ and the order parameter $\Delta$ is calculated as $\Delta = -g \sum_{\eta}u_{\eta}v_{\eta}^{*}$, in which $g$ is given by $1/k_f a = 8\pi E_f/g k_f^3 + \sqrt{4E_c / \pi^2 E_f}$~\cite{SandroReview}. Here $a$ is the 3D s-wave scattering length characterizing the interaction between atoms of different spins, while $E_f = \hbar^2 k_f^2 / 2m$ and $k_f = \left(3\pi^2n\right)^{1/3}$ are the Fermi energy and momentum of an ideal Fermi gas of density $n$. The cut-off energy $E_c$ is introduced in order to remove the ultraviolet divergences in the TDBdG equations with contact potentials. The density of the gas is $n \mbox(\textbf{r},t) = 2 \sum_{\eta} \left|v_{\eta} \mbox(\textbf{r},t)\right|^2$. Since the potential $U$ has no $y$ or $z$ dependence, we write the BdG eigenfunctions as $u_\eta \mbox(\textbf{r}) = u_\eta (x) e^{i (k_y y + k_z z)}$ and $v_\eta \mbox(\textbf{r}) = v_\eta (x) e^{i (k_y y + k_z z)}$, in which $k_y$ and $k_z$ are quantized according to $k_y = 2\pi \alpha_y / L_\bot$ and $k_z = 2\pi \alpha_z / L_\bot$, where $\alpha_y$ and $\alpha_z$ are integers and $L_\bot$ is the width of the box in the $y$- and $z$-directions.

As initial states at $t=0$, we find stationary solutions of Eq.~(\ref{eq:tdbdg}). This has been done to investigate stationary black solitons across the BEC-BCS crossover~\cite{Antezza}. We use the same technique to generate momentarily stationary solitons away from the trap center. When such an initial state is evolved in time, the black soliton is accelerated by the trap potential, and becomes grey. 

\begin{figure}[tbp]
\includegraphics[width=1.0\columnwidth]{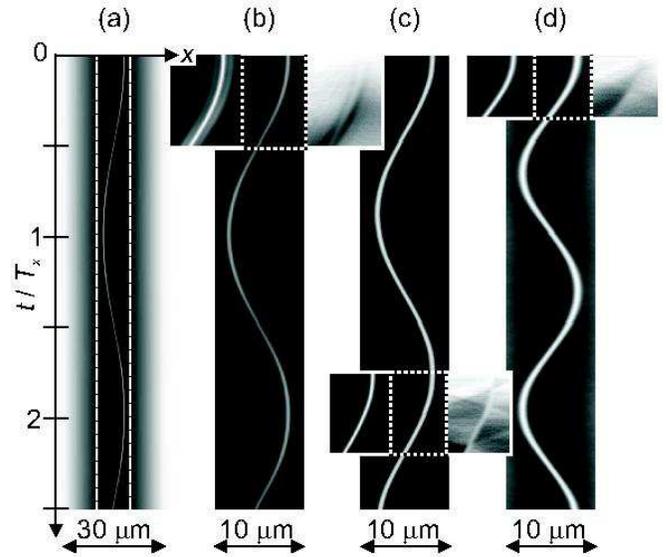} 
\caption{(a): Soliton oscillating in the density profile of a $^{40}K$ superfluid for $1/k_f a = -0.5$ with $\omega_x = 2 \pi 50$ rad s$^{-1}$, $L_\bot = 3.3$ $\mu$m and a peak density $n_p = 1.8 \times 10^{18}$ m$^{-3}$. (b): Enlargement of region contained within the dashed white box in (a). (c) \& (d): Corresponding enlargements for $1/k_f a = 0.0$ and $1.0$. Left and right insets in (b), (c) and (d) show the real part $\left|\Re\left(\Delta\right)\right|$ and imaginary part $\left|\Im\left(\Delta\right)\right|$ of the order parameter in the regions contained within the dotted white boxes. Horizontal bars show scale.}
\label{f1}
\end{figure}

Figure~\ref{f1} presents three typical simulations of the TDBdG equations. Panel (a) shows a soliton oscillating in the density profile of a $^{40}K$ superfluid for $1/k_f a = -0.5$, with $\omega_x = 2 \pi 50$ rad s$^{-1}$, $L_\bot = 3.3$ $\mu$m and a peak density $n_p = 1.8 \times 10^{18}$ m$^{-3} = n_{1d}(0)/L_\bot^2$. Panel (b) is an enlargement of the central region of panel (a) contained within the dashed white box. Panels (c) and (d) are equivalent enlargements for $1/k_f a = 0$ and $1$ respectively. We take $E_c = 30 E_f$ ($E_c = 50 E_f$) for $1/k_f a < 0$ ($1/k_f a \geq 0$). For $1/k_f a = -0.5$ [Figs.~\ref{f1}(a) and (b)], the soliton creates a shallow depression in the density of the cloud, on either side of which are smaller oscillations, known as Friedel oscillations. We also plot the real part $\left|\Re\left(\Delta\right)\right|$ and imaginary part $\left|\Im\left(\Delta\right)\right|$ of the order parameter in the region contained within the dotted white box in Fig.~\ref{f1}(b) as left and right insets respectively. Initially $\left|\Im\left(\Delta\right)\right|$ is zero, indicating that $J_\varphi = \pi$. As the soliton accelerates, the density depression becomes shallower, $\left|\Im\left(\Delta\right)\right|$ increases from zero and $J_\varphi$ reduces. The insets also show that both $\left|\Re\left(\Delta\right)\right|$ and $\left|\Im\left(\Delta\right)\right|$ contain Friedel oscillations in the vicinity of the soliton~\cite{brand}, as in the density profile. This is in contrast to solitonic solutions of the GP equation, which always have a constant imaginary component of the order parameter~\cite{levsandro2}. 

As $1/k_f a$ increases, the profile of the solitons in $n$ and $\Delta$ tends to that of GP solitons. At unitarity [Fig.~\ref{f1}(c)], the Friedel oscillations are barely visible in $n$, but $\left|\Im\left(\Delta\right)\right|$ (right inset) still contains a small dip at the position of the soliton. When $1/k_f a$ reaches $1.0$ [Fig.~\ref{f1}(d)], $\left|\Im\left(\Delta\right)\right|$ (right inset) is almost constant across the cloud. We also observe that the density depression becomes deeper. For $1/k_f a = 1.0$, the density minimum in the soliton is close to zero when it is stationary at the apex of an oscillation. 

\begin{figure}[tbp]
\includegraphics[width=1.0\columnwidth]{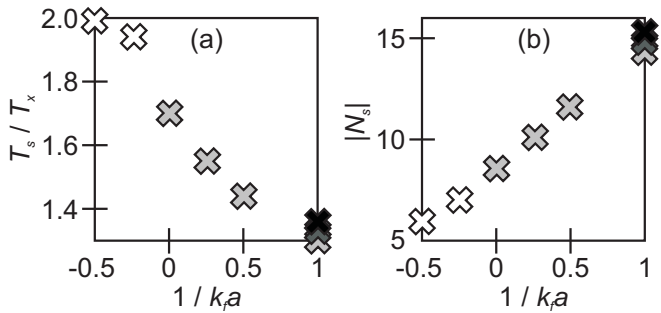} 
\caption{$T_s$ (a) and $N_s$ (b) plotted against $1/k_f a$. We find that $T_s = 1.7 T_x$ at unitarity. White (light gray, dark gray, black) crosses denote data for $E_c = 30 E_f$ ($50 E_f$, $75 E_f$, $100 E_f$). $N_s$ is calculated for transverse width $L_\bot = 3.3$ $\mu$m.}
\label{f2}
\end{figure}

Figure~\ref{f1} also illustrates that the period $T_s$ decreases as we move from the BCS to the BEC regime. This effect is quantified in Fig.~\ref{f2}(a), which plots $T_s$ against $1/k_f a$. The graph shows that $T_s$ drops rapidly as we move from the BCS to unitary regimes, before tending to the GP prediction of $\sqrt{2} T_x$~\cite{Busch2} in the BEC limit of large $1/k_f a$. Our prediction of $T_s = (1.7 \pm 0.05) T_x$ at unitarity is consistent with the value of $\sqrt{3} T_x$ given in Ref.~\cite{brand}. It is computationally difficult to reach convergence for large $1/k_f a$, because a large number of states must be included in order to describe the formation of Bosonic molecules. To illustrate the gradual convergence towards the GP prediction, we plot three points for $1/k_f a = 1$ with $E_c = 50 E_f$, $75 E_f$ and $100 E_f$, with a light gray, dark gray and black cross respectively. We also note that $T_s$ for $1/k_f a = -0.5$ is lower than expected by looking at the general trend. This is a real effect that occurs because the pair size is becoming comparable with the width of the cloud. It can be avoided by reducing $\omega_x$.

To compare the numerical results with the prediction of Eq.~(\ref{eq:final}), we must also calculate $N_s$ and $d J_\varphi / dV$. The quantity $N_s$ may be determined from stationary solutions of Eq.~(\ref{eq:tdbdg}). We plot results in Fig.~\ref{f2}(b) as a function of $1/ k_f a$. The graph shows that $N_s$ increases monotonically with $1/ k_f a$, reflecting that the soliton becomes much deeper while its width remains almost the same~\cite{Antezza}.

\begin{figure}[tbp]
\includegraphics[width=0.8\columnwidth]{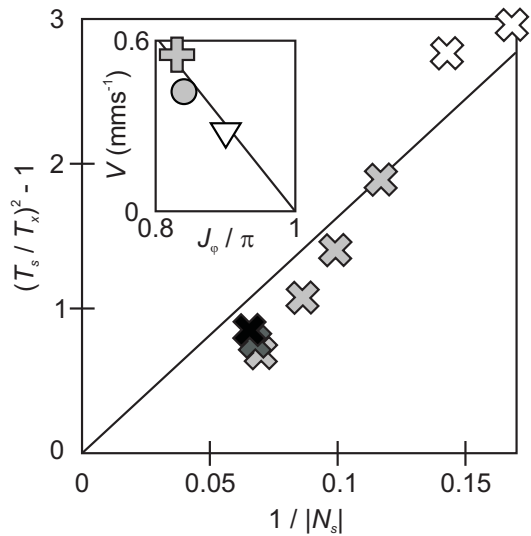} 
\caption{$\left[\left(T_s/T_x\right)^2 - 1\right]$ plotted against $1 / \left|N_s\right|$. White (light gray, dark gray, black) symbols denote data for $E_c = 30 E_f$ ($50 E_f$, $75 E_f$, $100 E_f$). Solid line shows the prediction of Eq.~(\ref{eq:final2}). Inset shows $V$ versus $J_\varphi$, with data points for $1/k_f a = -0.25$ (triangle), $0$ (circle) and $1.0$ (plus sign). Solid line shows the GP prediction for $a = 1/k_f$ [Eq.~(\ref{eq:gpsol})].}
\label{f2b}
\end{figure}

We determine $d J_\varphi / dV$ by measuring $V$ and $J_\varphi$ as the soliton passes the center of the trap. In the inset in Fig.~\ref{f2b}, we plot results for $1/k_f a = -0.25$, $0$ and $1.0$, with a triangle, circle and plus sign respectively. As expected, the result for $1/k_f a = 1$ lies close to the GP prediction~\cite{levsandro2} (black line) for $a = 1/k_f$ and small $V$, which is
\begin{equation}
V = \sqrt{\pi \hbar^2 n_p/4 k_f m^2} \left(\pi - J_\varphi\right) .
\label{eq:gpsol}
\end{equation}  
Suprisingly, the results for $1/k_f a = -0.25$ and $0$ also lie near the black line. The variation in $d J_\varphi / dV$ is comparable to the error in our simulations so, to a good approximation, we may take $d J_\varphi / dV$ to be a constant, given by Eq.~(\ref{eq:gpsol}). Given this and $m_B = 2m$, Eq.~(\ref{eq:final}) becomes
\begin{equation}
\left(T_s/T_x\right)^2 - 1 = - \left(3/\pi\right)^{1/6} L_\bot^2 n_p^{2/3}/N_s.
\label{eq:final2}
\end{equation}


In Fig.~\ref{f2b} we plot Eq.~(\ref{eq:final2}) (solid line) together with the numerical results (crosses); the two predictions agree within the accuracy of our TDBdG simulations. Figure~\ref{f2b} illustrates that the shallow solitons in the BCS regime accelerate slower than the deep solitons in the BEC regime. In the former case, $|N_s|$ is smaller, since more particles (unpaired fermions) are present inside the soliton. According to Eq.~(\ref{eq:final2}), this implies a larger $T_s$, in agreement with our numerical simulations. This is analagous to the increase in the period of dark solitons when they are filled by an impurity, creating bright-dark solitons~\cite{Busch,BeckerSH}.

\begin{figure}[tbp]
\includegraphics[width=1.0\columnwidth]{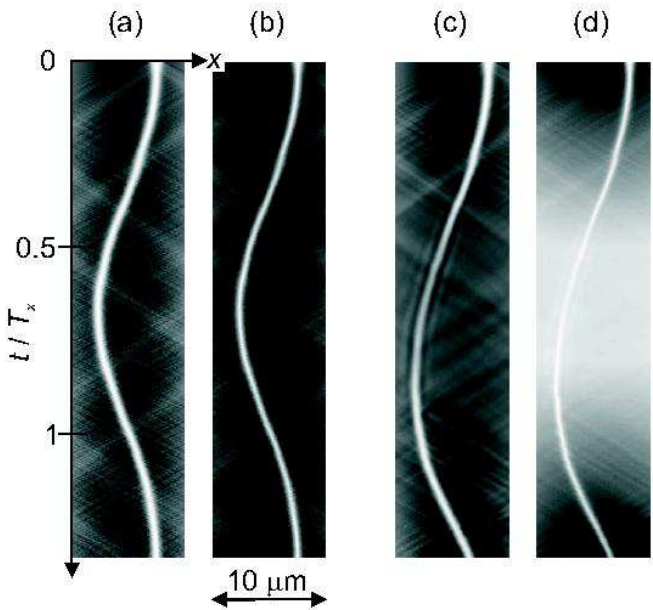} 
\caption{Proposed experimental scheme to detect the variation of $T_s$ with $1/k_f a$. Panels (a) [(c)] and (b) [(d)] show $n$ and $\left|\Delta\right|$ during protocol A [B]. Horizontal bar shows scale.}
\label{f3}
\end{figure}


{\it Experimental protocol.} We propose an experiment to observe the variation of $T_s$ across the BEC-BCS crossover. As in Fig.~\ref{f1}, we consider a $^{40}K$ superfluid with $\omega_x = 2 \pi 50$ rad s$^{-1}$, $L_\bot = 3.3$ $\mu$m and $n_p = 1.8 \times 10^{18}$ m$^{-3}$. However, in this case, the initial state does not contain a soliton. Instead, we create a soliton dynamically, using techniques that have been realised in experiment~\cite{BurgerSH, AndersonSH, DenschlagSH}. Firstly, we create a hole in the density of the initial state [Fig.~\ref{f3}(a)] by adding a narrow potential spike at $x = 2.5$ $\mu$m~\cite{Carr}. Secondly, at $t=0$ the cloud at $x<2.5$ ($x>2.5$) $\mu$m is imprinted with a phase of zero ($\pi$). The potential spike is then smoothly ramped down to zero over 0.5 ms ($= 0.025 T_x$) to minimise sound production. We refer to this procedure as protocol A. The procedure creates a black soliton plus some sound~\cite{Carr}. Since we form the soliton away from the center of the trap, it oscillates [Fig.~\ref{f3}(a)]. The amplitude of the order parameter is not significantly affected [Fig.~\ref{f3}(b)].

We now consider a second procedure, referred to as protocol B. The soliton is created in the same way but, from $t=0$ to 10 ms ($=0.5 T_x$), we ramp the scattering length until $1/k_f a = -0.5$. This reduces the speed and depth of the soliton, and Friedel oscillations appear in the density profile [Fig.~\ref{f3}(c)]. Also, the amplitude of the order parameter reduces dramatically [Fig.~\ref{f3}(d)]. Unfortunately, the soliton is now too shallow to be observed experimentally, so we ramp back to $1/k_f a = 1$ from $t=16$ ($=0.8 T_x$) to 26 ms ($=1.3 T_x$). The soliton now becomes deeper again, the Friedel oscillations disappear [Fig.~\ref{f3}(c)], and the order parameter increases [Fig.~\ref{f3}(d)]. By comparing the position of the soliton following protocol B to that following protocol A, the experimentalist may prove that $T_s$ has been increased. 

{\it Summary.} In this work, starting from very general assumptions, we derive universal relations valid for both Bosonic and Fermionic superfluids, relating the soliton energy and oscillation period $T_s$ to observable quantities. Then we solve the TDBdG equations to test these relations in a nontrivial theoretical model, and also to provide numerical predictions for $T_s$ in realistic conditions. Our TDBdG simulations, although exceeding challenging, represent also a new opportunity to probe unknown dynamics across the BEC-BCS crossover. We hope that our work, in particular our proposed protocol, will stimulate experiments to test our predictions.

We thank J. Brand and R. Liao for fruitful discussions and for sharing the preliminary results of their work~\cite{brand}. We thank G. Watanabe for helpful discussions. After this work was completed, we became aware of an independent study of grey solitons in the BEC-BCS crossover with stationary BdG equations~\cite{carrpaper}.

\bibliography{biblio}

\begin{thebibliography}{21}
\expandafter\ifx\csname natexlab\endcsname\relax\def\natexlab#1{#1}\fi
\expandafter\ifx\csname bibnamefont\endcsname\relax
  \def\bibnamefont#1{#1}\fi
\expandafter\ifx\csname bibfnamefont\endcsname\relax
  \def\bibfnamefont#1{#1}\fi
\expandafter\ifx\csname citenamefont\endcsname\relax
  \def\citenamefont#1{#1}\fi
\expandafter\ifx\csname url\endcsname\relax
  \def\url#1{\texttt{#1}}\fi
\expandafter\ifx\csname urlprefix\endcsname\relax\def\urlprefix{URL }\fi
\providecommand{\bibinfo}[2]{#2}
\providecommand{\eprint}[2][]{\url{#2}}

\bibitem[{\citenamefont{Kevrekidis et~al.}(2008)\citenamefont{Kevrekidis,
  Frantzeskakis, and Carretero-Gonz{\'{a}}lez}}]{kev}
\bibinfo{author}{\bibfnamefont{P.~G.} \bibnamefont{Kevrekidis}},
  \bibinfo{author}{\bibfnamefont{D.~J.} \bibnamefont{Frantzeskakis}},
  \bibnamefont{and}
  \bibinfo{author}{\bibfnamefont{R.}~\bibnamefont{Carretero-Gonz{\'{a}}lez}},
  \emph{\bibinfo{title}{Emergent Nonlinear Phenomena in Bose-Einstein
  Condensates, Theory and Experiment}} (\bibinfo{publisher}{Springer-Verlag},
  \bibinfo{year}{2008}).

\bibitem[{\citenamefont{Busch and Anglin}(2001)}]{Busch}
\bibinfo{author}{\bibfnamefont{T.}~\bibnamefont{Busch}} \bibnamefont{and}
  \bibinfo{author}{\bibfnamefont{J.~R.} \bibnamefont{Anglin}},
  \bibinfo{journal}{Phys. Rev. Lett.} \textbf{\bibinfo{volume}{87}},
  \bibinfo{pages}{010401} (\bibinfo{year}{2001}).

\bibitem[{\citenamefont{\mbox{C. Becker} \mbox{\textit{et
  al.}}}(2008)}]{BeckerSH}
\bibinfo{author}{\bibnamefont{\mbox{C. Becker} \mbox{\textit{et al.}}}},
  \bibinfo{journal}{Nature Physics} \textbf{\bibinfo{volume}{4}},
  \bibinfo{pages}{496} (\bibinfo{year}{2008}).

\bibitem[{\citenamefont{\mbox{M. W. Zwierlein} \mbox{\textit{et
  al.}}}(2005)}]{KetterleSH}
\bibinfo{author}{\bibnamefont{\mbox{M. W. Zwierlein} \mbox{\textit{et al.}}}},
  \bibinfo{journal}{Nature} \textbf{\bibinfo{volume}{435}},
  \bibinfo{pages}{1047} (\bibinfo{year}{2005}).

\bibitem[{\citenamefont{Antezza et~al.}(2007)\citenamefont{Antezza, Dalfovo,
  Pitaevskii, and Stringari}}]{Antezza}
\bibinfo{author}{\bibfnamefont{M.}~\bibnamefont{Antezza}},
  \bibinfo{author}{\bibfnamefont{F.}~\bibnamefont{Dalfovo}},
  \bibinfo{author}{\bibfnamefont{L.~P.} \bibnamefont{Pitaevskii}},
  \bibnamefont{and}
  \bibinfo{author}{\bibfnamefont{S.}~\bibnamefont{Stringari}},
  \bibinfo{journal}{Phys. Rev. A.} \textbf{\bibinfo{volume}{76}},
  \bibinfo{pages}{043610} (\bibinfo{year}{2007}).

\bibitem[{\citenamefont{Challis et~al.}(2007)\citenamefont{Challis, Ballagh,
  and Gardiner}}]{Challis}
\bibinfo{author}{\bibfnamefont{K.~J.} \bibnamefont{Challis}},
  \bibinfo{author}{\bibfnamefont{R.~J.} \bibnamefont{Ballagh}},
  \bibnamefont{and} \bibinfo{author}{\bibfnamefont{C.~W.}
  \bibnamefont{Gardiner}}, \bibinfo{journal}{Phys. Rev. Lett.}
  \textbf{\bibinfo{volume}{98}}, \bibinfo{pages}{093002}
  (\bibinfo{year}{2007}).

\bibitem[{\citenamefont{Fedichev et~al.}(1999)\citenamefont{Fedichev, Muryshev,
  and Shlyapnikov}}]{fedichev}
\bibinfo{author}{\bibfnamefont{P.~O.} \bibnamefont{Fedichev}},
  \bibinfo{author}{\bibfnamefont{A.~E.} \bibnamefont{Muryshev}},
  \bibnamefont{and} \bibinfo{author}{\bibfnamefont{G.~V.}
  \bibnamefont{Shlyapnikov}}, \bibinfo{journal}{Phys. Rev. A.}
  \textbf{\bibinfo{volume}{60}}, \bibinfo{pages}{3220} (\bibinfo{year}{1999}).

\bibitem[{\citenamefont{Konotop and Pitaevskii}(2004)}]{Lev}
\bibinfo{author}{\bibfnamefont{V.~V.} \bibnamefont{Konotop}} \bibnamefont{and}
  \bibinfo{author}{\bibfnamefont{L.}~\bibnamefont{Pitaevskii}},
  \bibinfo{journal}{Phys. Rev. Lett.} \textbf{\bibinfo{volume}{93}},
  \bibinfo{pages}{240403} (\bibinfo{year}{2004}).

\bibitem[{\citenamefont{\mbox{S. Burger} \mbox{\textit{et
  al.}}}(1999)}]{BurgerSH}
\bibinfo{author}{\bibnamefont{\mbox{S. Burger} \mbox{\textit{et al.}}}},
  \bibinfo{journal}{Phys. Rev. Lett.} \textbf{\bibinfo{volume}{83}},
  \bibinfo{pages}{5198} (\bibinfo{year}{1999}).

\bibitem[{\citenamefont{\mbox{B. P. Anderson} \mbox{\textit{et
  al.}}}(2001)}]{AndersonSH}
\bibinfo{author}{\bibnamefont{\mbox{B. P. Anderson} \mbox{\textit{et al.}}}},
  \bibinfo{journal}{Phys. Rev. Lett.} \textbf{\bibinfo{volume}{86}},
  \bibinfo{pages}{2926} (\bibinfo{year}{2001}).

\bibitem[{\citenamefont{\mbox{J. Denschlag} \mbox{\textit{et
  al.}}}(2000)}]{DenschlagSH}
\bibinfo{author}{\bibnamefont{\mbox{J. Denschlag} \mbox{\textit{et al.}}}},
  \bibinfo{journal}{Science} \textbf{\bibinfo{volume}{287}},
  \bibinfo{pages}{97} (\bibinfo{year}{2000}).

\bibitem[{\citenamefont{Ishikawa and Takayama}(1980)}]{ishikawa}
\bibinfo{author}{\bibfnamefont{M.}~\bibnamefont{Ishikawa}} \bibnamefont{and}
  \bibinfo{author}{\bibfnamefont{H.}~\bibnamefont{Takayama}},
  \bibinfo{journal}{J. Phys. Soc. Jpn.} \textbf{\bibinfo{volume}{49}},
  \bibinfo{pages}{1242} (\bibinfo{year}{1980}).

\bibitem[{\citenamefont{Shevchenko}(1988)}]{shev}
\bibinfo{author}{\bibfnamefont{S.}~\bibnamefont{Shevchenko}},
  \bibinfo{journal}{Sov. J. Low Temp. Phys.} \textbf{\bibinfo{volume}{14}},
  \bibinfo{pages}{553} (\bibinfo{year}{1988}).

\bibitem[{lie()}]{lieblin}
\bibinfo{note}{Interestingly, Eq.~(\ref{eq:PsPc}) implies that
  $\left|P_{c}\right| \leq \pi \hbar n_{1d0} m/m_B$. This property is also
  valid for the ``soliton-like'' branch of elementary exitations in a 1D Bose
  gas; see E. H. Lieb and W. Liniger, Phys. Rev. \textbf{130}, 1605 (1963).}

\bibitem[{foo()}]{foot2}
\bibinfo{note}{A formal derivation of Eq.~(\ref{Dphi}) will be given
  elsewhere.}

\bibitem[{\citenamefont{Giorgini et~al.}(2008)\citenamefont{Giorgini,
  Pitaevskii, and Stringari}}]{SandroReview}
\bibinfo{author}{\bibfnamefont{S.}~\bibnamefont{Giorgini}},
  \bibinfo{author}{\bibfnamefont{L.~P.} \bibnamefont{Pitaevskii}},
  \bibnamefont{and}
  \bibinfo{author}{\bibfnamefont{S.}~\bibnamefont{Stringari}},
  \bibinfo{journal}{Rev. Mod. Phys.} \textbf{\bibinfo{volume}{80}},
  \bibinfo{pages}{1215} (\bibinfo{year}{2008}).

\bibitem[{\citenamefont{Liao and Brand}(2010)}]{brand}
\bibinfo{author}{\bibfnamefont{R.}~\bibnamefont{Liao}} \bibnamefont{and}
  \bibinfo{author}{\bibfnamefont{J.}~\bibnamefont{Brand}},
  \bibinfo{journal}{e-print arXiv:1011.5337}  (\bibinfo{year}{2010}).

\bibitem[{lev()}]{levsandro2}
\bibinfo{note}{L. Pitaevskii and S. Stringari, \textit{Bose-Einstein
  Condensation} (Oxford University Press, 2003); C. J. Pethick and H. Smith,
  \textit{Bose-Einstein Condensation in dilute gases} (Cambridge University
  Press, 2008).}

\bibitem[{\citenamefont{Busch and Anglin}(2000)}]{Busch2}
\bibinfo{author}{\bibfnamefont{T.}~\bibnamefont{Busch}} \bibnamefont{and}
  \bibinfo{author}{\bibfnamefont{J.~R.} \bibnamefont{Anglin}},
  \bibinfo{journal}{Phys. Rev. Lett.} \textbf{\bibinfo{volume}{84}},
  \bibinfo{pages}{2298} (\bibinfo{year}{2000}).

\bibitem[{\citenamefont{Carr et~al.}(2001)\citenamefont{Carr, Brand, Burger,
  and Sanpera}}]{Carr}
\bibinfo{author}{\bibfnamefont{L.~D.} \bibnamefont{Carr}},
  \bibinfo{author}{\bibfnamefont{J.}~\bibnamefont{Brand}},
  \bibinfo{author}{\bibfnamefont{S.}~\bibnamefont{Burger}}, \bibnamefont{and}
  \bibinfo{author}{\bibfnamefont{A.}~\bibnamefont{Sanpera}},
  \bibinfo{journal}{Phys. Rev. A.} \textbf{\bibinfo{volume}{63}},
  \bibinfo{pages}{051601(R)} (\bibinfo{year}{2001}).

\bibitem[{\citenamefont{Spuntarelli et~al.}(2010)\citenamefont{Spuntarelli,
  Carr, Pieri, and Strinati}}]{carrpaper}
\bibinfo{author}{\bibfnamefont{A.}~\bibnamefont{Spuntarelli}},
  \bibinfo{author}{\bibfnamefont{L.~D.} \bibnamefont{Carr}},
  \bibinfo{author}{\bibfnamefont{P.}~\bibnamefont{Pieri}}, \bibnamefont{and}
  \bibinfo{author}{\bibfnamefont{G.~C.} \bibnamefont{Strinati}},
  \bibinfo{journal}{e-print arXiv:1011.4257}  (\bibinfo{year}{2010}).

\end{thebibliography}

\end{document}